\documentclass{PoS}

\usepackage{xspace}
\usepackage{relsize}
\usepackage{subfigure}

\def\Journal#1#2#3#4{{#1} {\bf #2}, #3 (#4)}

\def\PLB{{\em Phys. Lett.}  B}
\def\PRL{\em Phys. Rev. Lett.}
\def\PRD{{\em Phys. Rev.} D}

\def\etal {{\em{at al.}}\xspace}

\def\ks{\ensuremath{K^0_\mathrm{S}}}
\def\kl{\ensuremath{K^0_\mathrm{L}}}

\def\be{\begin{equation}}
\def\ee{\end{equation}}
\def\bea{\begin{eqnarray}}
\def\eea{\end{eqnarray}}

\def\iab{\ensuremath{\mathrm{ab}^{-1}}\xspace}
\def\ifb{\ensuremath{\mathrm{fb}^{-1}}\xspace}
\def\ipb{\ensuremath{\mathrm{pb}^{-1}}\xspace}
\def\bfactory{\ensuremath{B} Factory\xspace}
\def\bfactories{\ensuremath{B} Factories\xspace}

\def\CP{\ensuremath{\mathrm{CP}}\xspace}
\def\B{\ensuremath{B}\xspace}
\def\Bz{\ensuremath{B^0}\xspace}

\def\K{\ensuremath{K}\xspace}
\def\Kz{\ensuremath{K^0}\xspace}

\def\D{\ensuremath{D}\xspace}
\def\Dz{\ensuremath{D^0}\xspace}

\def\jpsi{\ensuremath{J/\psi}\xspace}

\def\sintwobeta{\ensuremath{\sin 2 \beta}\xspace}
\def\sintwobetaeff{\ensuremath{\sin 2 \beta_{\mathrm eff}}\xspace}
\def\deltasm{\ensuremath{\Delta_{\mathrm{SM}}\xspace}}

\def\tanbeta{\ensuremath{\tan \beta}\xspace}
\def\mhiggsp{\ensuremath{m_{H^+}}\xspace}

\def\rk{\ensuremath{R_{\mathrm{K}}}}
\def\br{\ensuremath{{\cal{B}}\xspace}}

\def\FourS{\ensuremath{\Upsilon(\mathrm{4S})}}

\newcommand\vud {\ensuremath{V_{\mathrm{ud}}}\xspace}
\newcommand\vus {\ensuremath{V_{\mathrm{us}}}\xspace}
\newcommand\vub {\ensuremath{V_{\mathrm{ub}}}\xspace}

\newcommand\vcb {\ensuremath{V_{\mathrm{cb}}}\xspace}

\def\stat{\mathrm{(stat.)}}
\def\syst{\mathrm{(syst.)}}

\newcommand{\e}      [1]   {{\ensuremath{\times 10^{{#1}}}}}
\newcommand{\su}      [1]   {{\ensuremath{SU({#1})}}}
\newcommand\url[1] {{\tt #1}}

\def\fL{\ensuremath{f_{\mathrm{L}}}\xspace}
\def\afb{\ensuremath{A_{\mathrm{FB}}}\xspace}
\def\ai{\ensuremath{A_{\mathrm{I}}}\xspace}

\def\babar{\mbox{\slshape B\kern-0.1em{\smaller A}\kern-0.1em
    B\kern-0.1em{\smaller A\kern-0.2em R}}\xspace}
\def\belle{Belle\xspace}

\title{Flavour Physics at B-factories and other machines}

\ShortTitle{Flavour Physics at B-factories and other machines}

\author{\speaker{A.~J.~Bevan}\\
        Department of Physics, Queen Mary University of London, Mile End Road, London E1 4NS, England\\
        E-mail: \email{a.j.bevan@qmul.ac.uk}}

\abstract{ I review some of the highlights of results from non-hadron collider flavor experiments shown at the EPS conference in the summer of 2009. These include highlights of the latest results from the \babar\ and \belle B-factories, CLEO-c, BES-III, NA48, KTeV, KLOE, NA62, MEG, and $\mu \to e$ conversion experiments. }

\FullConference{The 2009 Europhysics Conference on High Energy Physics,\\
		 July 16 - 22 2009\\
		 Krakow, Poland}

\begin{document}

\section{Introduction}

The highlight of flavor physics over the past year has been the accolade awarded to
Kobayashi and Maskawa by the Nobel Prize committee for their work on extending the 
theory of 2 generation quark mixing to include a third generation, and in doing so to
naturally allow for the phenomenon of \CP\ violation~\cite{bevan:nobelprize} via a 
single complex phase.  
This results in the Cabibbo-Kobayashi-Maskawa (CKM) quark-mixing mechanism~\cite{bevan:cabibbo,bevan:km}. 
After 37 years of testing, all observed \CP\ conserving and violating phenomenon in the 
quark sector have been found to agree with the CKM theory.  Much of these proceedings are 
devoted to experimental tests of predictions made by the CKM matrix.

The experimental results from non-hadron flavor experiments are summarized in these proceedings,
including results from the \babar\ and \belle B-factories, CLEO-c, BES-III, NA48, KTeV, KLOE, NA62,
MEG, and $\mu \to e$ conversion experiments.  Many of these measurements can be used to constrain
our knowledge of \CP violation and quark mixing, and in doing so they confirm that the CKM mechanism
is the leading order description of nature.  However the current precision of results allows for
significant new physics contributions as a second order correction to the CKM picture.  The remainder
of these proceedings discuss recent results from experiment.

\section{The B-factories}

\subsection{\boldmath The Unitarity Triangle}

The angles of the unitarity triangle (weak phases) are $\alpha$, $\beta$, and $\gamma$.  Some
of the literature, including journal papers from the \belle experiment, use an
alternate notation where $(\beta, \alpha, \gamma) = (\phi_1, \phi_2, \phi_3)$.
Only two of these angles are independent parameters, predicted by the CKM 
mechanism, the third is constrained via $\alpha+\beta+\gamma=180^\circ$.

\subsubsection{\boldmath The angle $\beta$}

The measurement of $\beta$ using $\Bz \to \jpsi \ks$ decays was the the raison d'etre 
of the \bfactories.  Not only has this measurement been made, but a number of ancillary 
measurements of $\B$ meson decays to final state comprising a Charmonium and neutral $K$ or $K^*$
have been made by these experiments.  One of the impressive achievements of both experiments
is the measurement of \sintwobeta from tree level processes with a Charmonium in the final state,
and the good agreement between all of these measurements.  Table~\ref{bevan:tbl:sintwobeta}
summarises these measurements in terms of $S$ and $C$, where $S=\sqrt{1-C^2}\sintwobeta$ in the SM.
Even though the \bfactories have surpassed their design luminosities and recorded some 1.5\iab
of data, the measurement of \sintwobeta is still statistically limited: $\beta = (21.1\pm 0.9)^\circ$~\cite{bevan:hfag}.

\begin{table}[!h]
\caption{Measurements of \sintwobeta made from tree level Charmonium decays at the \bfactories.  The combined results quote statistical and systematic uncertainties added in quadrature.  Other results quote statistical errors followed by statistical uncertainties, and where only one error is given it is the statistical error.}\label{bevan:tbl:sintwobeta}
\begin{center}
\begin{tabular}{l|ccc}\hline
Mode            &  \babar  & \belle & Average \\ \hline\hline 
$\jpsi\ks$      & $0.657\pm 0.036\pm 0.012$ & $0.643 \pm 0.038$ & $-$ \\
$\jpsi\kl$      & $0.694\pm 0.061\pm 0.031$ & $0.641\pm 0.057$ & $-$\\
$\jpsi K^{*0}$  & $0.601\pm 0.239\pm 0.087$ & $-$ & $-$ \\
$\eta_c\ks$     & $0.925\pm 0.160\pm 0.057$ & $-$ & $-$\\
$\chi_{1c}\ks$  & $0.614\pm 0.160\pm 0.040$ & $-$ & $-$ \\
$\psi(2S)\ks$   & $0.897\pm 0.100\pm 0.036$ & $0.718\pm 0.090\pm 0.031$ & $0.798\pm 0.071$\\
combined        & $0.691\pm 0.031$ & $0.650\pm 0.034$& $0.672\pm 0.023$ \\ \hline
\end{tabular}
\end{center}
\end{table}

\subsubsection{\boldmath The angle $\alpha$}

The second angle to be measured by the \bfactories is $\alpha$.  Measurements use \B decays
into $u\overline{u}d$ final states including $\pi\pi$, $\rho\pi$, $\rho\rho$, and $a_1\pi$.
The tree level contribution to this process competes with a significant loop (penguin) 
amplitude that has a different weak phase to the tree.  As a result one has to determine 
or constrain the bias from penguin amplitudes in order to measure $\alpha$.  There are 
a number of different recipes in the literature describing the steps required to constrain
penguin contributions and thus measure $\alpha$.  The most popular method is the 
Gronau-London \su{2} Isospin analysis of $\pi\pi$ and 
$\rho\rho$ decays~\cite{bevan:alpha:gronaulondon}, although it is also possible to use an 
\su{3} based approach to measure $\alpha$ from $\rho^+\rho^-$ and $K^{*0}\rho^+$ 
decays~\cite{bevan:alpha:beneke} to obtain a measurement of $\alpha$ with similar precision.

The most precise constraint on $\alpha$ comes from $\B \to \rho\rho$ decays, where two 
improvements have occurred in the past year: i) a proof-of principle time-dependent CP
asymmetry measurement has been performed in $\Bz\to\rho^0\rho^0$ decays, and ii) an
updated branching fraction measurement of $\B^+\to\rho^+\rho^0$ has been performed.
A precision time-dependent measurement of $\Bz\to\rho^0\rho^0$ at a future Super Flavour
Factory would help us remove ambiguities in the measurement of $\alpha$, coming from the 
construct used to constrain penguin contributions, and in taking the arcsine of 
$S/\sqrt{1-C^2}$.  The main impact on the constraint on $\alpha$ comes from the updated
branching fraction measurement of $\B^+\to\rho^+\rho^0$.  This is important as the 
$\rho^+\rho^0$ amplitude normalizes the base of the Isospin triangles used to constrain
penguin contributions.  The latest world average branching fraction of $(24\pm 2)\e{-6}$
means that the base of the triangle is a similar length to the side given by $\rho^+\rho^-$.
The corollary of this is that uncertainties in penguin contributions are all degenerate,
and the precision on $\alpha$ using this methods is significantly reduced.  
The precision on $\alpha$ from $\rho\rho$ decays is now $\pm 6.5^\circ$ with 
the \su{2} approach~\cite{bevan:rrz}, and $\pm 7^\circ$ with the \su{3} 
approach~\cite{bevan:rr}.

\begin{figure}[!h]
\begin{center}
  \resizebox{8.0cm}{!}{\includegraphics{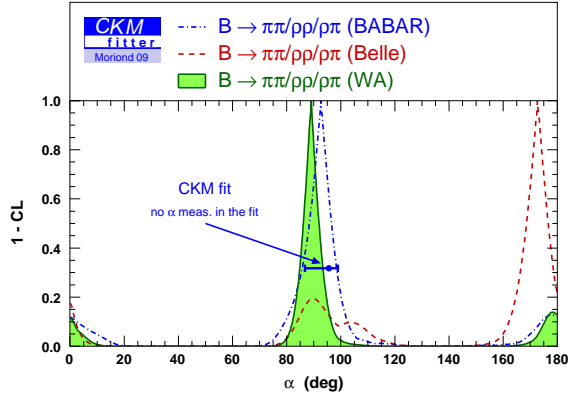}}
\end{center}
 \caption{The distribution of one minus the confidence level for $\alpha$ obtained
   from $\pi\pi$, $\rho\pi$, and $\rho\rho$ decays (From CKM Fitter~\cite{bevan:ckmfitter}).}
\label{bevan:fig:alpha}
\end{figure}

The world average constraint on $\alpha$ from $\pi\pi$, $\rho\pi$ and $\rho\rho$ decays
is $(89^{+4.4}_{-4.2})^\circ$ (CKM Fitter)~\cite{bevan:ckmfitter} and 
$(92\pm 7)^\circ$ (UTfit)~\cite{bevan:utfit}.  The constraints from CKM Fitter
are shown in Fig.~\ref{bevan:fig:alpha} and do not include a measurement using 
$B\to a_1^\pm\pi^\mp$ decays with a precision of $13^\circ$
on $\alpha$.  At the end of data taking for the \bfactories it is interesting to note
that of the four measurements of $\alpha$, the one with the worst precision 
$\B\to \pi\pi$ was originally envisaged as the best way to measure this angle.

\subsubsection{\boldmath The angle $\gamma$}

The third unitarity triangle angle to measure is $\gamma$.  This is relatively simple to 
measure from a theoretical viewpoint, however it is experimentally challenging
by virtue of the low decay rates of interesting channels.  $\gamma$ is measured from 
charged \B meson decays to $D^{(*)}K^{*}$ final states.

There are several theoretical schemes used to constrain this angle on the market.  The most
popular ones are the ADS~\cite{bevan:ads}, GLW~\cite{bevan:glw} and GGSZ~\cite{bevan:ggsz} 
methods.  In the longer term a precision measurement will be obtained from each of these
methods, however given the current data samples, it is the latter method that dominates the 
precision of our knowledge on $\gamma$.  This method utilizes the interference pattern 
in the $D\to K_S\pi^+\pi^-$ Dalitz plot of $\B\to \overline{D}^{(*)}K^{*}$ decays, where the 
difference between the $\B^+$ and $B^-$ decays contains information about this weak phase.  
Using 605\ifb of data \belle have been able to determine 
$\gamma = (78.4^{+10.8}_{-11.6} \pm 3.6 \pm 8.9)^\circ$~\cite{bevan:gamma:belle}.
This last uncertainty is dominated by the lack of knowledge of the 
$D\to K_S\pi^+\pi^-$ Dalitz plot.  This error can be reduced as discussed later
in Section~\ref{bevan:sec:cleo}.  \babar have performed similar 
measurements~\cite{bevan:gamma:babar}, where
the precision of the final result differs as \babar use less data, and they extract
a smaller value for the ratio of Cabibbo suppressed to allowed decays: $r_B$.

The world average constraint on $\gamma$ from the ADS, GLW and GGSZ methods
is $(70^{+27}_{-30})^\circ$ (CKM Fitter)~\cite{bevan:ckmfitter} and 
$(78\pm 12)^\circ$ (UTfit)~\cite{bevan:utfit} (See Fig.~\ref{bevan:fig:gamma}).  

\begin{figure}[!h]
\begin{center}
  \resizebox{8.0cm}{!}{\includegraphics{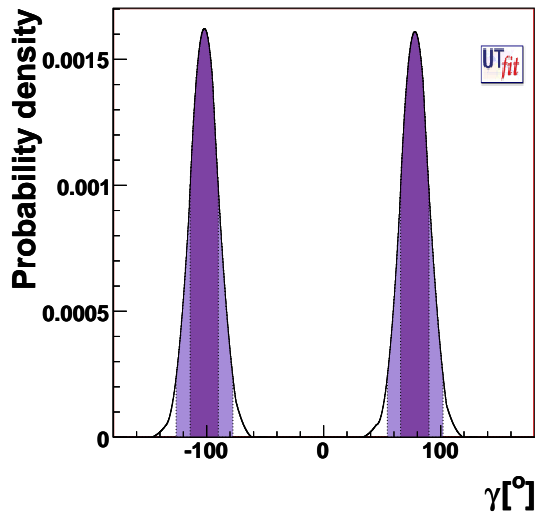}}
\end{center}
 \caption{The distribution of the probability for $\gamma$ obtained
   from the ADS, GLW and GGSZ methods (From UTFit~\cite{bevan:utfit}).}
\label{bevan:fig:gamma}
\end{figure}

\subsubsection{\boldmath Summarizing the angles and testing the standard model}
\label{sec:gamma}

The direct measurements of $\alpha$ and $\beta$ alone are sufficient
to constrain the unitarity triangle to a precision of $5^\circ$.  This
constitutes a precision test of the CKM picture of CP violation in \B\
meson decays.  If one includes $\gamma$ it is possible to check that
the angles of the unitarity triangle sum to $180^\circ$ as expected.
The precision of these tests is $(180^{+27}_{-30})^\circ$ and $(191 \pm 14)^\circ$
for results from the CKM fitter~\cite{bevan:ckmfitter} and
UTfit~\cite{bevan:utfit} groups, respectively.

Having constrained the unitarity triangle, it is possible to start to test
the SM description in various ways.  One of these tests has been to compare
the measured value of \sintwobeta\ in Charmonium decays with the measurements
of \sintwobetaeff\ made in loop (penguin) dominated processes.  The caveat to
making such a comparison is that the penguin dominated modes may have additional
topologies that could lead to a difference between \sintwobetaeff\ and \sintwobeta.
If these SM corrections \deltasm\ are well known then any residual difference
$\Delta S =  \sintwobetaeff - \sintwobeta -\deltasm$ would be from new physics.
It is recently been pointed out that there are additional tests one should make,
by using indirect constraints on CKM related processes in order to compute
the expected value of \sintwobeta\ from the SM.  This inferred value of
\sintwobeta\ should be compared with both the directly measured Charmonium and
penguin dominated measurements.  Figure~\ref{bevan:fig:deltas_measurements} summarizes the
different constraints on \sintwobeta, where the difference between the measured and
inferred values of \sintwobeta have a significance of 2.1 to $2.7\sigma$.

\begin{figure}[!h]
\begin{center}
  \resizebox{10.0cm}{!}{
       \includegraphics{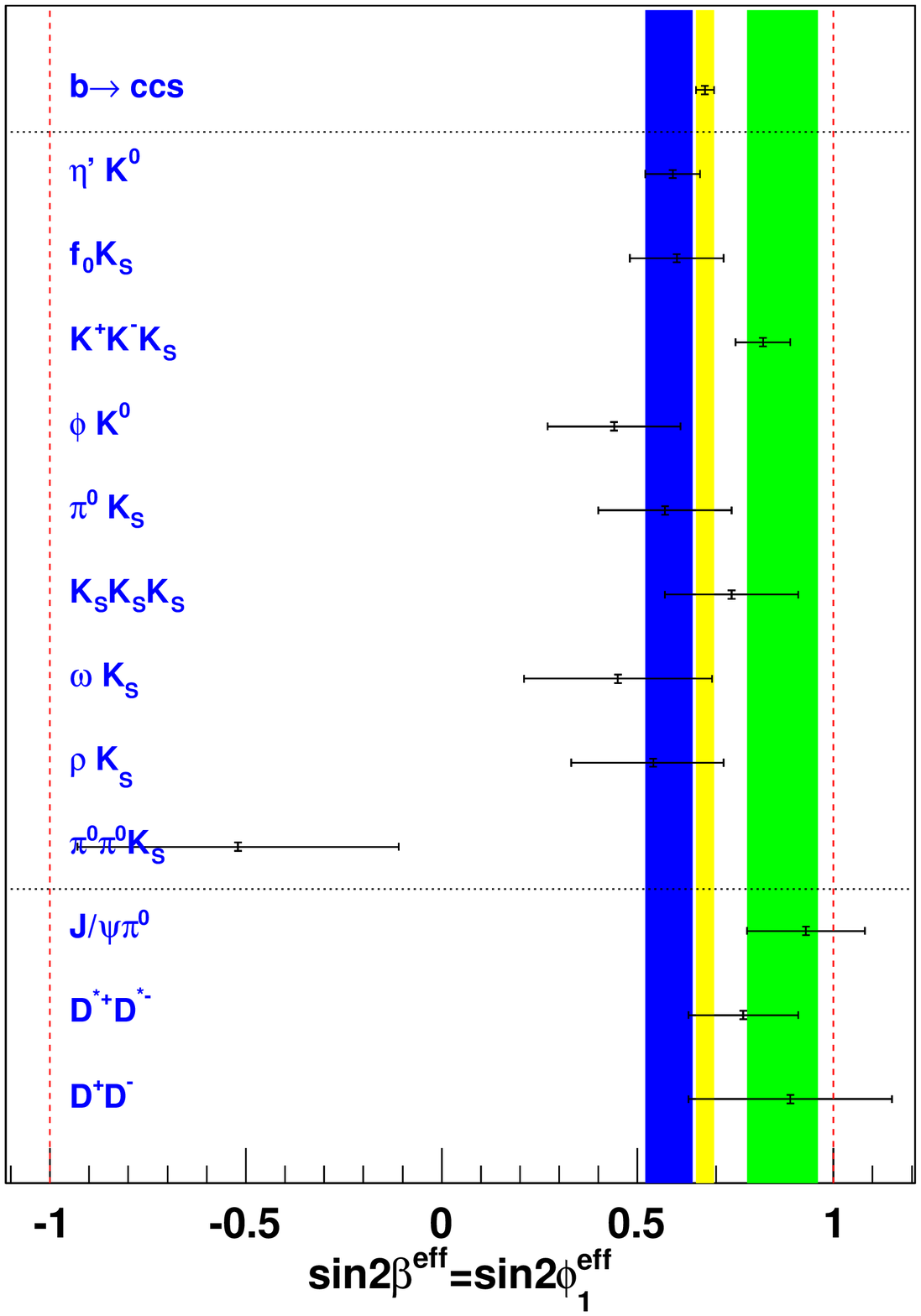}
  }
  \resizebox{16.0cm}{!}{
       \includegraphics{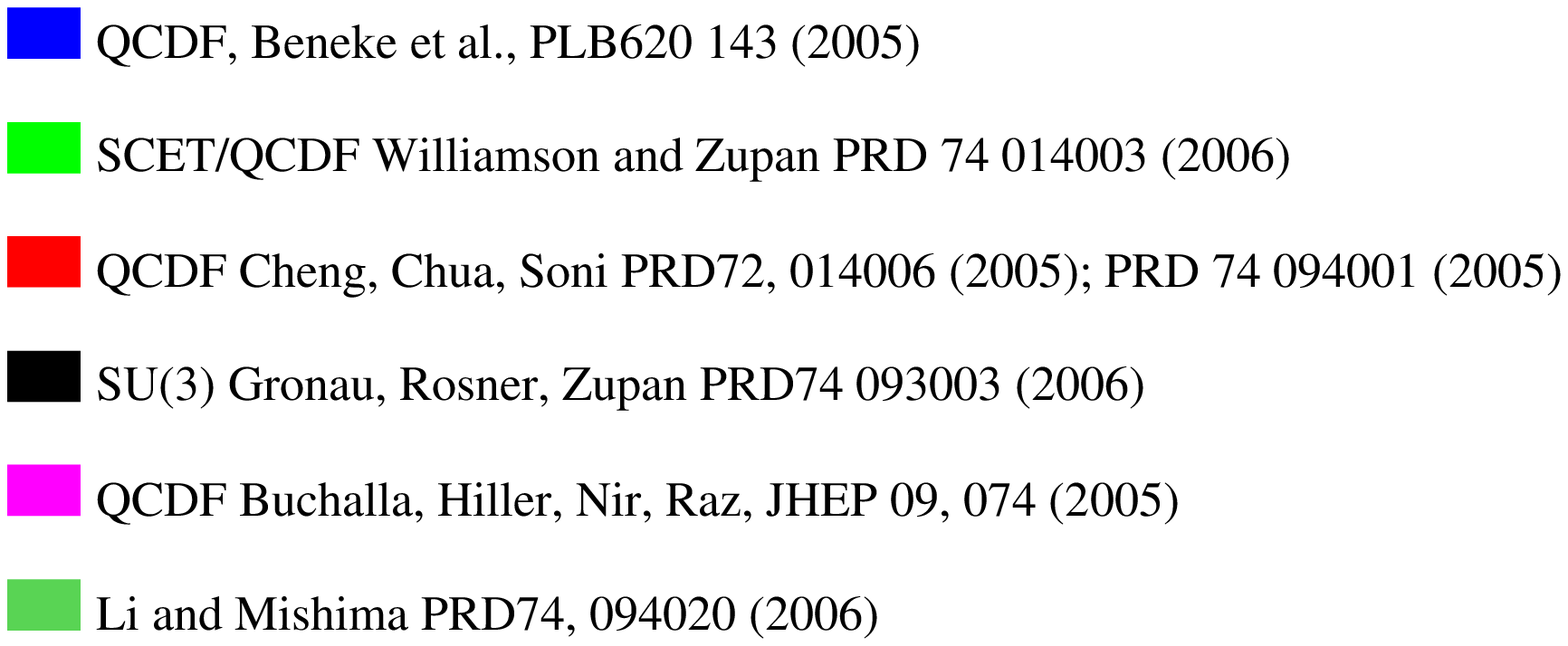}
       \includegraphics{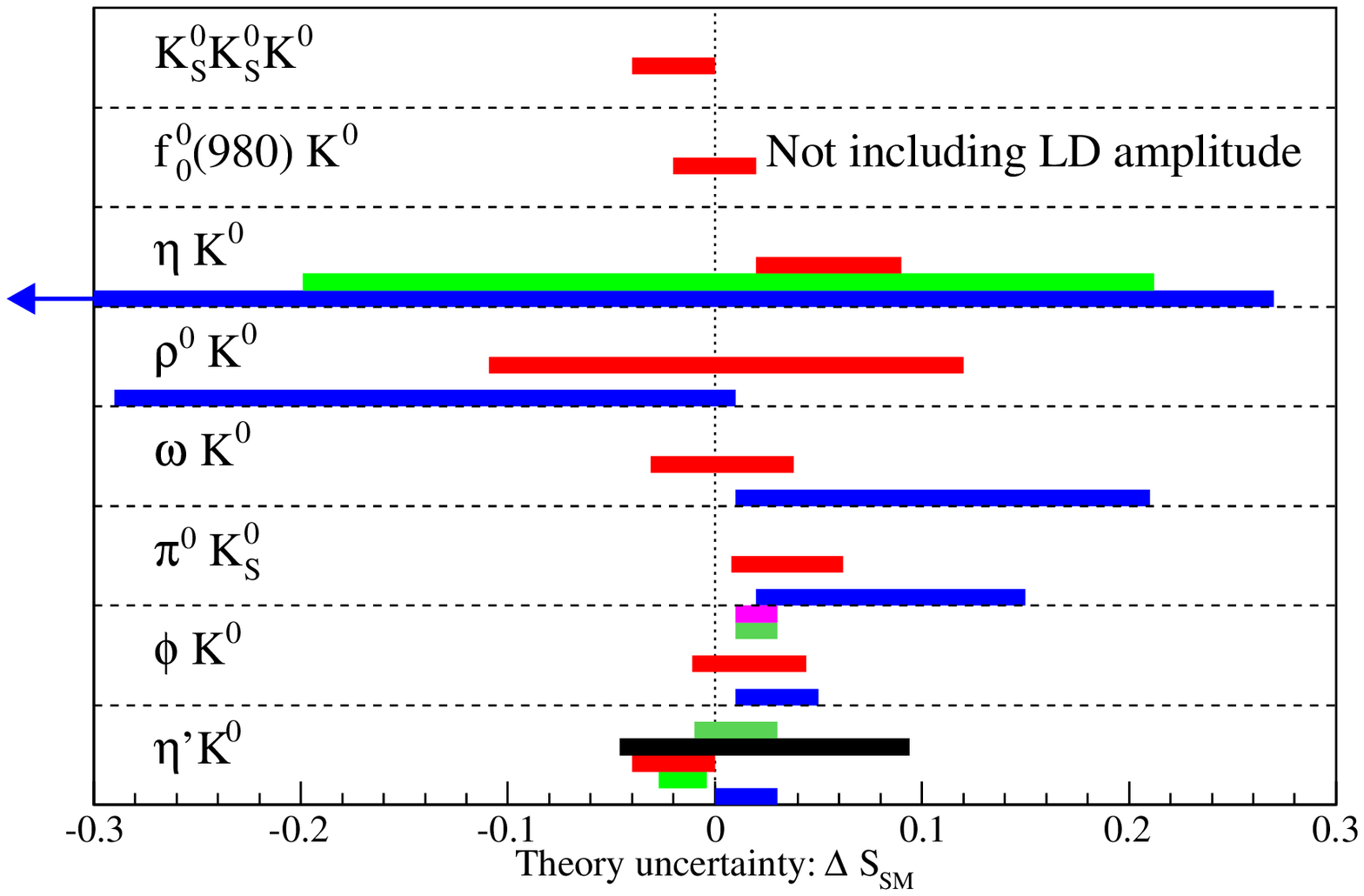} 
  }
\end{center}
 \caption{(top) The measured values of \sintwobeta\ in (yellow) Charmonium decays,
   (blue) penguin decays, and (green) inferred from indirect measurements.
   The constraint from penguin decays is an average of the two theoretically
   clean modes: $\Bz \to \eta'K^0$ and $\Bz \to \phi K^0$.  (bottom) Theoretical
   uncertainties \deltasm\ calculated 
   for $b\to s$ penguin decays, with references given in the legend on the left.}
\label{bevan:fig:deltas_measurements}
\end{figure}

\subsection{Direct CP violation}

The B-factories have performed many searches for direct \CP violation over the 
past decade.  These searches have resulted in the observation ($>5\sigma$ significance) 
of this phenomenon in two decay modes: $\Bz \to K^\pm \pi^\mp$ and $\Bz\to\pi^+\pi^-$, 
and a handful of channels indicating evidence ($>3\sigma$) of this effect including
$B$ decays to $\eta K^*$, $\eta K^\pm$, $\rho^0 K^\pm$, $\rho^\pm\pi^\mp$, and 
$D^{0*}K^\pm$ final states.  Theoretically the level of direct \CP\ violation depends
on strong phase differences between interfering amplitudes.  These phase differences
are hard to calculate so it is difficult to try and interpret these measurements.

One conundrum that has been perplexing the community for several years is the 
so called $K\pi$ puzzle.  In the SM the difference between the direct \CP asymmetry
in $\Bz \to K^\pm \pi^\mp$ and $\B^\pm \to K^\pm \pi^0$ is expected to be small
and positive.  The world average of this quantity turns out to be $-0.148\pm 0.028$,
which is clearly different from expectations.  It has been noted that the difference 
observed here could be a sign of new physics, however the question of weather a more 
detailed understanding of the hadronic dynamics of these decays would resolve the 
discrepancy remained a possibility.  At this conference S. Mishima presented work
that was able to account for the large negative difference in 
asymmetries~\cite{bevan:directcpv:mishima}.

\subsection{The polarization puzzle}

The study of \B meson decays to final states with two vector ($J^P=1^-$) particles V has
presented us with a decade long puzzle.  The fraction of longitudinally polarized events \fL\
in such decays is expected to be $1-m^2_V/m^2_b \sim 1$.  This naive expectation works
well for some decays such as the $\B\to \rho\rho$ channels discussed above.  However there
are clear deviations from this expectation, most notably in $B\to \phi K^{*}$ decays where
$\fL\sim 0.5$.  
Experimentally it is possible that
deviations from the expectation of $\fL=1$ could come from new physics, although additional
experimental data would be useful in resolving this conundrum.
The experimental situation is summarized in Figure~\ref{bevan:fig:btovv}, 
where it is clear that precision measurements of a number of these rare Charmless \B\ decays
would help to elucidate the non-trivial pattern of \fL.  

In order to try and address the limited data, experimentalists have been searching
for rare \B\ decays to final states that also include axial-vector ($J^P=1^+$) mesons 
A.  Searches for the AV final states $a_1^\pm\rho^\mp$~\cite{bevan:polarization:a1rho},
 $a_1^\pm K^{*\mp}$, and 
combinations of a $b_1$ particle with a $\rho$ or $K^*$\cite{bevan:polarization:b1v} have so far yielded negative
results.  However \babar\ recently reported observation of the decay 
$\Bz\to a_1^\pm a_1^\mp$ with a branching fraction central value of 
$(11.8 \pm 2.6 \pm 1.6) \e{-6}$.  The corresponding measurement of 
$\fL= 0.31 \pm 0.24$~\cite{bevan:polarization:a1a1} which is another non-trivial
result on the polarization puzzle.
In the longer term one could envisage obtaining 
information on the unitarity triangle angle $\alpha$ from this channel.  

\begin{figure}[!h]
\begin{center}
  \resizebox{12.0cm}{!}{
       \includegraphics{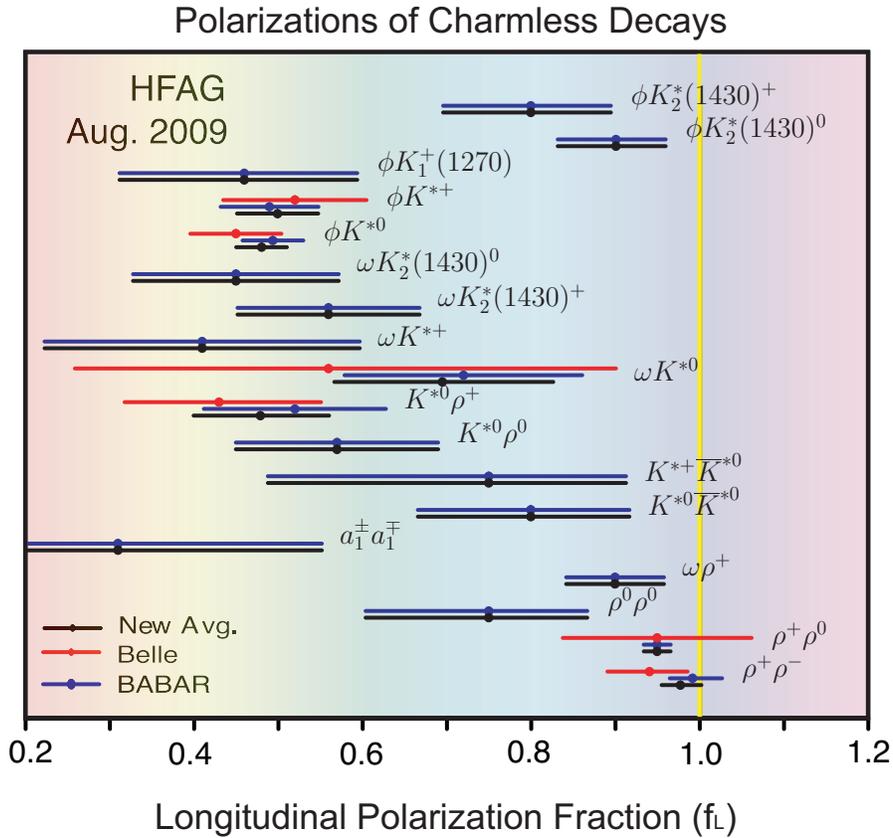}
}
\end{center}
 \caption{The measured values of \fL from \B\ meson decays to final states with two spin one particles (Figure from Ref.~\cite{bevan:hfag}).}
\label{bevan:fig:btovv}
\end{figure}

\subsection{Rare B decays}

The \bfactory experiments are able to constrain new physics through the study of rare or SM suppressed
processes.  I highlight only two of the many different channels that can be studied: $B^\pm \to \tau^\pm\nu$,
and $B\to s\ell^+\ell^-$.

\subsection{\boldmath $B^\pm \to \tau^\pm\nu$}

Identifying this final state is experimentally challenging as there are two neutrinos from the signal decay.
The method used by the \bfactories is to fully reconstruct one of the $B$ mesons in the event, and then
fully reconstruct the signal $B$, modulo the neutrinos.  If a signal event has been correctly identified
at this stage, then there will be no other activity in the detector.  Following this reconstruction scheme,
it is possible to use the remaining energy in the calorimeter as a discriminating variable which goes into
a multi-dimensional maximum-likelihood fit in order to extract signal.  Both \babar\ and \belle\ follow this
scheme, and in doing so they obtain a branching fraction world average of $(1.73\pm 0.35)\e{-4}$ for this decay.
The SM prediction is $(1.6\pm 0.4)\e{-4}$, where the dominant uncertainties on this prediction come from
the uncertainty in \vub\ and $f_B$.  

Having made a measurement of this branching fraction, it is possible to constrain NP parameter space for 
a number of different models, including 2HDM~\cite{bevan:raredecays:2hdm}, MSSM~\cite{bevan:raredecays:mssm}, 
and unparticles~\cite{bevan:raredecays:unparticles}.  If we take the 
example of 2HDM, the observed branching fraction could be enhanced or suppressed by a factor of 
$(1-m_B^2 \tan^2\beta/m_H^2)^2$.  The corresponding constraint on $m_H^+$ as a function of $\tan\beta$ exceeds 1TeV for $\tan\beta > 40$, and is about 100 GeV for $\tan\beta = 5$.  The constraint on $m_H^+$ vs. 
$\tan\beta$ obtained for MSSM is similar.  The constraint expected to be obtained from 
the LHC is typically $\leq 160 GeV$ (95\% C.L.) using 30\ifb of 
data~\cite{bevan:lhcmssm}.  The $gg/gb \to t[b]H^+$ channel, with $H^+\to \tau^+ \nu$ 
can provide more stringent exclusions for large values of $\tan\beta$, however it is 
worth noting that the constraint already obtained indirectly from rare $B$ 
decays provides stronger constraints than expected for a direct search using 
a 30\ifb data sample at the LHC.

\subsection{\boldmath $B \to s\ell^+\ell^-$}

In analogy to the $B$ meson decays to final states with two vector particles that are discussed above,
there are many interesting observables that can be measured in the decay $B$ mesons to inclusive and 
exclusive $s\ell^+\ell^-$ final states, where $s$ is a strangeness one meson.  These observables include
\fL, forward-backward asymmetry \afb, Isospin asymmetry \ai, and the ratio of rates to $e^+e^-$ and 
$\mu^+\mu^-$ final states $R_{s\ell\ell}$.  Recent results from the \bfactories
show that data are consistent with expectations from the SM, however it is clear that there are limited
statistics available~\cite{bevan:sll:babar,bevan:sll:belle}.  If one compares the data to the \afb expectation (see Fig.~\ref{bevan:fig:raredecays:sll}) of a new physics scenario
with a sign-flipped Wilson coefficient of the effective Hamiltonian $C_7^{\mathrm{eff}}$, then it is clear
that the data prefer this expectation to that of the SM.  Any significant quantitative
test of this agreement will require more statistics than currently available, so this 
tantalising hint will either be refuted or confirmed by future experiments.

\begin{figure}[!ht]
\begin{center}
\resizebox{8cm}{!}{
\includegraphics{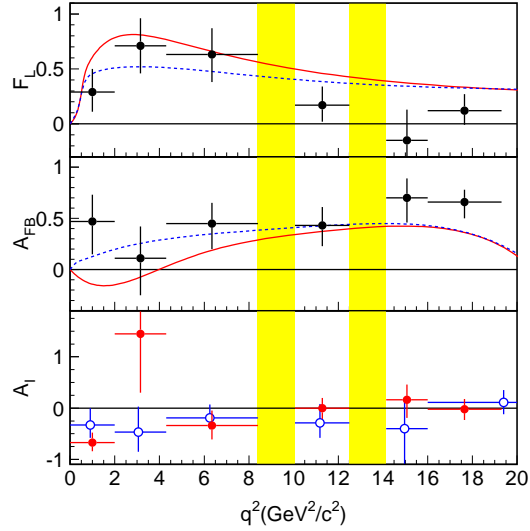}
}
\end{center}
\caption{The forward-backward asymmetry distributions from 
Belle~\cite{bevan:sll:belle}. The solid line corresponds to the SM expectation,
and the dashed line corresponds to a NP scenario with a sign-flipped $C_7^{\mathrm{eff}}$.}
\label{bevan:fig:raredecays:sll}
\end{figure}

\subsection{\boldmath Lepton flavour violation in $\tau$ decays}

The \bfactory experiments are also $\tau$ factories.  There is a wide range of SM tests
and NP searches that can be performed at these experiments.  One of the most promising
sets of LFV measurements that can be performed are the $\tau\to 3\ell$ decays.  These
are expected to occur in the SM with $\br\sim 10^{-54}$~\cite{bevan:taulfv:smratelll} based on the
known level of neutrino mixing.

The \bfactories have searched for all possible three-lepton final states using blind analyses optimised on
signal and background Monte Carlo simulated data, as well as data sidebands.  The upper limits obtained from
these analyses are summarised in Table~\ref{bevan:tbl:taulfv}.  The sensitivity of the current experiments 
reaches down to $1 - 2 \e{-8}$.  This is sufficient to constrain a number of LFV NP scenarios, and also 
surpasses the sensitivity expectations of the LHC~\cite{bevan:lhclfv}.

\begin{table}[!h]
\caption{90\% C.L. Upper limits on the branching fraction of $\tau \to 3 \ell$ decays.}\label{bevan:tbl:taulfv}
\begin{center}
\begin{tabular}{l|ccc}\hline
Mode & $\epsilon$ [\%] \babar (Belle) & UL [\e{-8}] \babar (Belle) \\ \hline \hline
$e^+e^-e^+$       & 8.6  (6.0)   & 2.9 (2.7)\\
$e^+e^-\mu^+$     & 8.8  (9.3)   & 2.2 (1.8)\\
$e^+e^+\mu^-$     & 12.6 (11.5)  & 1.8 (1.5)\\
$e^+\mu^-\mu^+$   & 6.4  (6.1)   & 3.2 (2.7)\\
$e^-\mu^+\mu^+$   & 10.2 (10.1)  & 2.6 (1.7)\\
$\mu^+\mu^-\mu^+$ & 6.6  (7.6)   & 3.3 (2.1)\\  \hline
\end{tabular}
\end{center}
\end{table}

There are a number of other searches for LFV in $\tau$ decay and many measurements of SM processes that 
have been performed at the \bfactories.  Unfortunately time constraints did not permit these to be discussed.

\subsection{\boldmath $|\vub|$}

In the past 12 months there has been little change in this area.  The multiple inclusive measurements
of $\vub$ are all in agreement with each other.  Similarly the multiple exclusive measurements all
agree.  There is a tension between the inclusive and exclusive results, that remains at the level of 
$1 - 2 \sigma$ as shown in Figure~\ref{bevan:fig:vub}.  The measured value of \sintwobeta\ is more compatible
with the exclusive $|\vub|$ measurement.

\begin{figure}[!ht]
\begin{center}
\resizebox{8cm}{!}{
\includegraphics{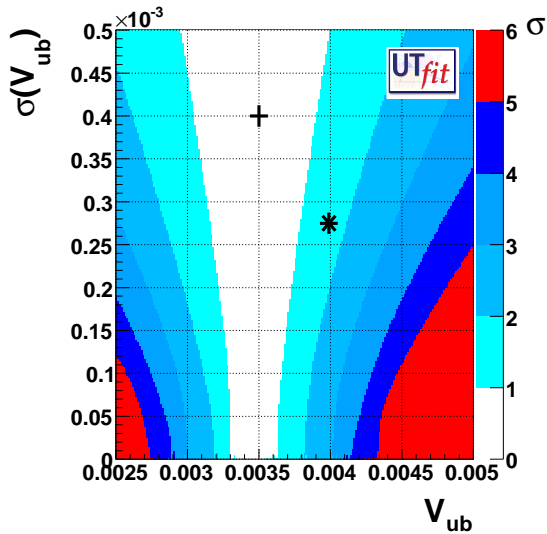}
}
\end{center}
\caption{The compatibility plot for the inclusive (*) and exclusive (+) $|\vub|$ measurements (From UTfit~\cite{bevan:utfit}).}
\label{bevan:fig:vub}
\end{figure}

\subsection{\boldmath $|\vcb|$}

As with the $|\vub|$ measurements, there is some disagreement between the inclusive and exclusive 
$|\vcb|$ determinations using $b\to c \ell \nu$ decays with $D^*$ mesons in the final state.  
These results are discussed in more detail in Refs.~\cite{bevan:vcb}.

\subsection{Charm mixing}

The charm production cross section at the \FourS\ is larger than the \B\ production cross section.
As a result the \B\ factories have recorded hundreds of millions of charm decays.  There is a very 
broad charm physics programme in studying this data, however both the focus and highlight of the 
past two years has been the study of neutral charm meson mixing.   

We can define mixing parameters $x$ and $y$ which are related to the mass and width differences 
($\Delta m$ and $\Delta\Gamma$) of
the two mass eigenstates $|D_1\rangle$ and $|D_2\rangle$.  Where the mass eigen-states are related
to the strong eigenstates via $p|D^0 \rangle\pm q|\overline{D}^0 \rangle$. The mixing parameter 
$x=\Delta m/\Gamma$, and $y=\Delta \Gamma / 2\Gamma$, where $\Gamma$ is the sum of the widths of 
$D_1$ and $D_2$.  There have been a number of different measurements of $x$ and $y$ performed
by the \bfactories, CLEO-c and CDF.  The combined average of these measurements has a significance
of a non-zero value of $x$ and $y$ which is $10.2\sigma$.  This constraint is shown in 
Figure~\ref{bevan:fig:charmmixing}.  It is still interesting to note that while charm mixing has been
firmly established, no single measurement has yet produced a signal greater than $5\sigma$.

\begin{figure}[!ht]
\begin{center}
\resizebox{10cm}{!}{
\includegraphics{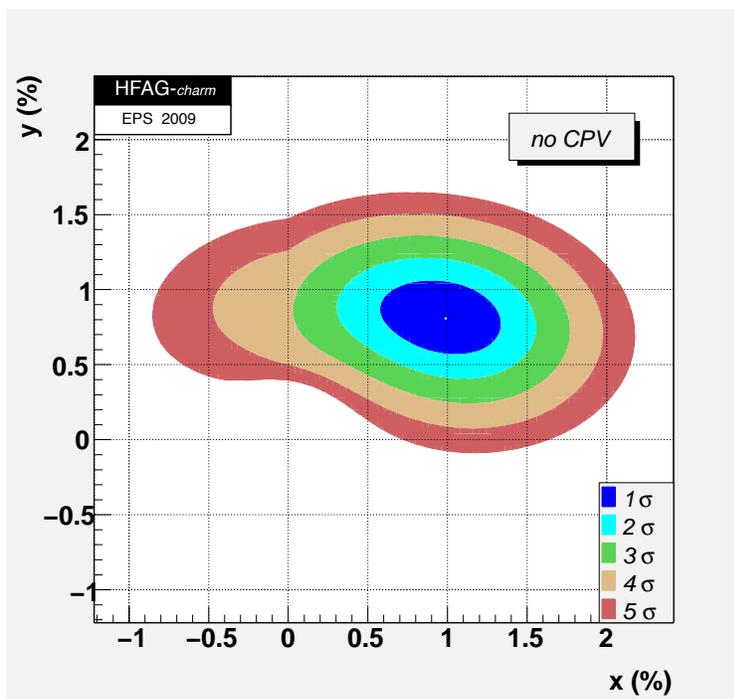}
}
\end{center}
\caption{The constraint on $x$ and $y$ from all available measurements given by HFAG~\cite{bevan:hfag}.  
         The significance of $(x, y) \neq (0, 0)$ is $10.2\sigma$.}
\label{bevan:fig:charmmixing}
\end{figure}

Now that charm mixing has been established, the next logical question is weather or not
there is \CP violation in charm decays, or mixing.  As one expects a small level of \CP violation 
in charm decays from SM related effects, and large effects (at the few \% level) 
would be a clear indication of new physics.  The search for \CP violation in mixing
yields a result compatible with \CP conservation and $|q/p|=1$.

Another possible test for charm mixing is the comparison of the lifetime measured for 
Cabibbo allowed $D\to K\pi$ and Cabibbo suppressed $D\to h^+h^-$ decays.  If these lifetimes
differ significantly, then mixing will have occurred.  This is tested via the measurement
of the quantity $y_{\CP} = \tau_{D\to K\pi} / \tau_{D\to h^+h^-}-1$, where a non-zero result 
indicates mixing.  The world average result for $y_{\CP} = (1.07\pm 0.21)\%$, which again 
confirms the effect of mixing in charm decays.

\subsection{Spectroscopy}
One of the surprises of the \bfactories was the many new states that have been uncovered
since 2003.  These started with Belle's discovery of the X(3872)~\cite{bevan:x3872}, and
a second boost to this activity was initiated by the discovery of the Y(4260) in the 
study of $J/\psi \pi\pi$ using ISR data at \babar~\cite{bevan:y4260}.  A total of 11 new Charmonium states
have been found as a result of these searches.  These fit into the pattern of expected
SM states above the open charm threshold.  However there are a number of details that remain 
unanswered about the nature of these states.  These questions range from simple confirmation of 
a state by a second experiment, as is the case for the recently discovered $Z(4430)$, to 
the nature of the particle itself, weather they are mesons, or some exotic hybrid state.
Given that the \bfactories are at the end of their data taking life, it is likely that 
many of these questions will remain unanswered until new experiments (such as the Super Flavour
Factories discussed below) are built.

\subsection{Light Higgs Searches}

It is possible to search for light scalar Higgs particles, such as the $A^0$ predicted by
NMSSM to have a mass $< 10 GeV/c^2$ using data collected at the $\Upsilon(3S)$ by \babar.
This mass range is inaccessible to LEP, the Tevatron, and the LHC.  If NMSSM is a more 
precise description of nature than the SM, we need to have low energy $e^+e^-$ collider 
programme to detect the $A^0$ and study it's properties.  \babar have recently reported the 
results of searches for $\Upsilon(3S) \to \gamma A^0$, with the $A^0$ decaying into either
a $\tau^+\tau^-$~\cite{bevan:lighthiggs:tautau}, $\mu^+\mu^-$~\cite{bevan:lighthiggs:mumu}, 
or a $\nu\overline{\nu}$~\cite{bevan:lighthiggs:nunu} final state.  For $m_{A^0}<10 GeV/c^2$
the $\tau^+\tau^-$ channel is expected to dominate.  No signal was found for any of the final
states studied, and the limits placed on $\Upsilon(3S) \to \gamma A^0$ are significant improvements
over previous studies.  The limits for the $\tau^+\tau^-$ channel are of the order of a few
$\times 10^{-5}$.

\subsection{Light Dark Matter Searches}

It is possible that light dark matter could be detected in decays of light mesons into 
invisible final states.  Here the amplitude of a meson decay into a pair of dark matter
particles could swamp the SM amplitude, leading to a significant enhancement in the 
measured branching fraction~\cite{bevan:lightdm:th}.  The \bfactories have searched
for $\Upsilon(1S)\to invisible$ final states in order to test this model. The \babar 
result uses $\Upsilon(3S)\to \Upsilon(1S)\pi\pi$, and tags
the $1S$ sample using two soft pions.  The limit placed on $\Upsilon(1S)\to invisible$
by \babar is $<3.0 \times 10^{-4}$ (90\% C.L.)~\cite{bevan:lightdm:babar}, 
which places non trivial constraints
on NP predictions, and is a factor of ten better than the previous 
search~\cite{bevan:lightdm:belle}.

\section{Super Flavour Factories}

The \bfactories have surpassed all expectations and accumulated 1.5\iab of data.  The type of physics that 
has been done at these experiments encompasses much more than was originally envisaged when they were 
being constructed.  The next phase in flavour physics studies using $e^+e^-$ colliders operating with 
a center of mass energy in the vicinity of the $\Upsilon(4S)$ anticipates a data sample of 50 to 75\iab
at a so-called Super Flavour Factory.  There are two proposals being pursued: Belle-II at KEK in 
Japan~\cite{bevan:belle2}, and SuperB near Frascati in Italy~\cite{bevan:cdr,bevan:valencia}.  The 
former aims to integrate 50\iab by the end of 2020, and the latter
experiment aims to integrate 75\iab of data on a similar time-scale.  The physics goals of these experiments
are to elucidate the nature of high-energy interactions through the study of rare or forbidden decays.  
This is possible through the precision study of rare decays, utilizing the uncertainty principle so
that virtual massive particles interfere with or dominate the SM amplitudes.
Such measurements have an energy reach way beyond that of the brute force approach taken by hadron colliders
such as the Tevatron and the LHC.  Until recently the KEK Super Flavour Factory intended to construct 
an accelerator based on a high current scheme.  However recently this has been abandoned in favor of 
the low emittance scheme developed for the SuperB programme in Italy.  In addition to the higher 
luminosity goals of SuperB, that experiment will be built with polarized electron beams and the ability
to collect data at the charm threshold center of mass energy corresponding to the $\psi(3770)$.
These additional features give the SuperB experiment a broader physics programme than Belle-II.

\section{CLEO-c}\label{bevan:sec:cleo}

\subsection{\boldmath Constraining the model uncertainty on $\gamma$}

As discussed in Section~\ref{sec:gamma}, there is a non-trivial model uncertainty 
on the measurement of $\gamma$ that comes from a lack of detailed understanding 
of the $D\to K_S\pi\pi$ Dalitz structure. If this model uncertainty were to have
remained, then the GGSZ method for measuring $\gamma$ would be limited to an ultimate
precision of the order of ten degrees.  However it is possible to utilize quantum
correlations at charm threshold to improve the knowledge of the Dalitz model, and 
hence the precision on $\gamma$~\cite{bevan:giri}.  CLEO-c have accumulated 
818\ipb of data running 
at a center of mass energy corresponding to charm threshold: $\psi(3770)$.
Using this data, CLEO-c have performed a detailed analysis of $D\to K_s h^+h^-$ decays, 
where $h=\pi, K$ ~\cite{bevan:cleocgamma}.  These results are estimated to reduce the 
model uncertainty on $\gamma$ from the GGSZ method from $9^\circ$ to $1.7^\circ$. As the
results are not systematically limited, measurements of this Dalitz plot in the future by 
BES-III and at SuperB could further reduce the model uncertainty on the GGSZ $\gamma$ 
measurement.

\subsection{Other charm results}

In addition to the aforementioned measurements, CLEO-c has accumulated 586\ipb of data above
charm threshold at a center of mass energy of 4170 MeV.  In addition to producing 
$D^0$ and $D^+$ mesons, one also produces $D_s$ particles.  A vast array of measurements
of the branching fractions and properties of $D^0$, $D^+$, and $D_s$ decays have been 
produced using these data~\cite{bevan:cleocbfs}.  Many of the $D_s$ results have been published using 300\ipb of data
in Ref.~\cite{bevan:cleocds}.  In addition to the aforementioned results, CLEO-c have 
managed to measure semi-leptonic $D$ decays~\cite{bevan:cleoc:semileptonic}, however time did not permit
these results to be discussed.

\section{Charmonium Factories: BES-III and KEDR}

The BES-III Charmonium factory has been running routinely, having achieved luminosities
of $3\times 10^{32} cm^{-1} s^{-2}$. Thus far the experiment has recorded 
$10^8$ $e^+e^- \to \psi(2S)$ transitions and is in the process of analyzing this data sample.
The detector is now well understood and has started to produce preliminary results.  One 
such result is the confirmation of the decay $\psi(2S)\to \pi^0 h_c$, with the subsequent
decay $h_c \to \gamma \eta_c$~\cite{bevan:bes3}.  BES-III is currently running on the 
$J/\psi$ resonance, and aims to integrate $3$ to $5\times 10^{8}$ decays at this center
of mass energy.  Later plans for this experiment include running at charm threshold in 
order to accumulate ${\cal O}(20 \ifb)$.

The KEDR experiment has performed precision measurements of the $J/\psi$ leptonic
width obtaining $\Gamma_{ee}^2 / \Gamma = 0.3355 \pm 0.0064 \pm 0.0048$ keV, which 
shows a marked improvement in precision relative to the DASP experiment performed 
in 1979~\cite{bevan:kedr}. The cross-section of $e^+e^- \to \mu^+\mu^-$ decays has also been 
accurately measured in the vicinity of the $J/\psi$.  In addition to these,
KEDR have produced precision measurements of the charged and neutral $D$ meson 
mass, resulting in the most precise measurement of $m_{D^+} = 1869.32 \pm 0.48 \pm 0.22$ 
MeV~\cite{bevan:kedr}.

\section{Kaon physics}
%
%
\subsection{\boldmath $\epsilon^\prime/\epsilon$}

A decade ago the NA48 and KTeV collaborations produced their first results
on the measurement of $\epsilon^\prime/\epsilon$ which is obtained via
the double ratio of \kl\ and \ks\ decays into pairs of neutral and charged
pions through
\begin{eqnarray}
R = 1 - 6\Re\left(\epsilon^\prime/\epsilon\right) = \frac{ N(\kl\to\pi^0\pi^0) / N(\ks\to\pi^0\pi^0) }{ N(\kl\to\pi^+\pi^-) / N(\ks \to\pi^+\pi^-)}
\end{eqnarray}
The results from these two experiments clearly indicated that 
$\epsilon^\prime/\epsilon\neq 0$, and in doing so established the phenomenon 
of direct CP violation in the SM.  During the last few years KTeV have
been working on finalizing systematic uncertainties related to calorimeter
cluster reconstruction and to acceptance effects in their analysis, and 
have now produced their final result.  The measurements of $\epsilon^\prime/\epsilon$
from these experiments are~\cite{bevan:epove}:
\begin{eqnarray}
 (\epsilon^\prime/\epsilon)_{\mathrm{NA48}} &=& (14.2 \pm 1.4 \stat \pm 1.7 \syst)\e{-4}\\
 (\epsilon^\prime/\epsilon)_{\mathrm{KTeV}} &=& (19.2\pm 1.1 \stat \pm 1.8 \syst)\e{-4}
\end{eqnarray}
and the corresponding world average is $(16.8\pm 1.4)\e{-4}$.

\subsection{\boldmath $|V_{\mathrm{us}}|$}

The quantity $|\vus|$ is a fundamental parameter of the SM.  The unitarity of the CKM matrix means
that we expect $|\vud|^2+|\vus|^2+|\vub|^2=1$, where for simplicity one can neglect the tiny 
contribution from \vub.  There has been some tension highlighted between the SM expectation of
unitarity and recent measurements \vus and \vud.  KLOE have updated their measurement of \vus,
through the study of $K\to \pi \ell \nu$ decays.  This has resulted in a set of measurements 
of $f_+(0) |\vus|$, where $f_+(0)$ is a form factor calculable on the lattice.  The KLOE 
result for $f_+(0) |\vus| = 0.2157 \pm 0.0006$~\cite{bevan:kloe:vus}.  
Using $f_+(0) = 0.9644 \pm 0.0049$ from UKQCD/RBC~\cite{bevan:lattice:vus}, and 
$\vud= 0.97418\pm 0.00025$ from $0^+ \to 0^+$ $\beta$ decays~\cite{bevan:vud}, the world 
average value of $\vus= 0.2237 \pm 0.0013$, which includes the aforementioned KLOE result.  
Using this result, it is clear that the measurements of 
\vus and \vud are compatible with unitarity.

\subsection{\boldmath $R_{\mathrm{K}}$}

One powerful test for new physics through lepton universality violation is the measurement of $R_K$, which is the ratio of branching ratios for charged kaons decaying into a $K_{e2}$ and $K_{\mu 2}$ final state (i.e. $e\nu$, and $\mu\nu$):
\begin{eqnarray}
R_K = \frac{\Gamma(K^\pm \to e^\pm \nu)}{\Gamma(K^\pm \to \mu^\pm \nu) }.
\end{eqnarray}
Enhancements from new physics can be parameterized in a number of ways, for example
Ref.~\cite{bevan:petronzio} uses $\Delta R = R_K^{measured}/R_K^{SM} = 1 + \Delta R_{NP}$
to search for signs of new physics giving deviation from the expected SM value $R_K^{SM}$. 
In order to compute the SM expectation of $R_K$ one requires a detailed knowledge of the 
effect of final state radiation for the electron mode~\cite{bevan:rk:radcorr}.  Given that this is 
theoretically understood, the crux of this NP search lies in the measurement of the ratio.  In itself,
the $R_K$ measurement is experimentally very challenging and limited by the experimental
determination of the rate of $K_{e2}$ decays.  The KLOE experiment at DA$\Phi$NE has accumulated
a data sample of 13,800 $K_{e2}$ decays.  Using this KLOE have achieved a 1\% precision measurement 
of $R_K= (2.493\pm 0.025\pm 0.019)\e{-5}$ which is still limited by statistics~\cite{bevan:kloe:rk}.  
Recently the NA62 
collaboration has released a preliminary result based on 40\% of its total data sample.  Using that
data with a sample of 51,089 $K_{e2}$ decays NA62 has reached a 0.6\% measurement of 
$R_K= (2.500\pm 0.012\pm 0.011)\e{-5}$~\cite{bevan:na62:rk}.  Here the reported systematic uncertainty is limited
by effects that will reduce with statistics.  The ultimate precision of $R_K$ from NA62 is expected 
to be 0.3\%.  The NA62 result dominates the world average of $R_K$.

\subsection{\boldmath $K\to \pi\nu\overline{\nu}$}

In a recently approved proposal NA62~\cite{bevan:na62}, will be modified in order to be able to perform 
a measurement of the branching fraction of $K^+\to \pi^+\nu\overline{\nu}$ with a precision
of 10\%.  This branching fraction provides a theoretically clean constraint on the apex of the
unitarity triangle.  In order to achieve this a number of new detector sub-systems are in
the process of being redesigned.  This includes a silicon pixel tracker whose purpose
is to tag the presence of a $K^+$ candidate in the detector; a new set of veto anti-counters
to surround the decay volume; straw trackers for reconstructing the trajectories of charged particles;
and a RICH based particle identification system.

The proposed KOTO~\cite{bevan:koto} experiment at JPARC aims to perform a 15\% measurement of the branching fraction
of the neutral mode $\kl\to \pi^0\nu\overline{\nu}$.  This is a theoretically clean measurement of 
the height of the unitarity triangle. 

Together with the measurement of $\epsilon_K$, it would be possible to put a competitive
constraint on the apex of the unitarity triangle from kaon decays.  Such a constraint could be
be used as a test of the compatibility of results from \B\ and $K$ decays in the 
quark mixing sector and \CP\ violation.  Figure~\ref{bevan:fig:kaonckm} illustrates a prediction
of the type of constraint on the unitarity triangle that would be possible from a combination 
of these three measurements.

\begin{figure}[!ht]
\begin{center}
\resizebox{10cm}{!}{
\includegraphics{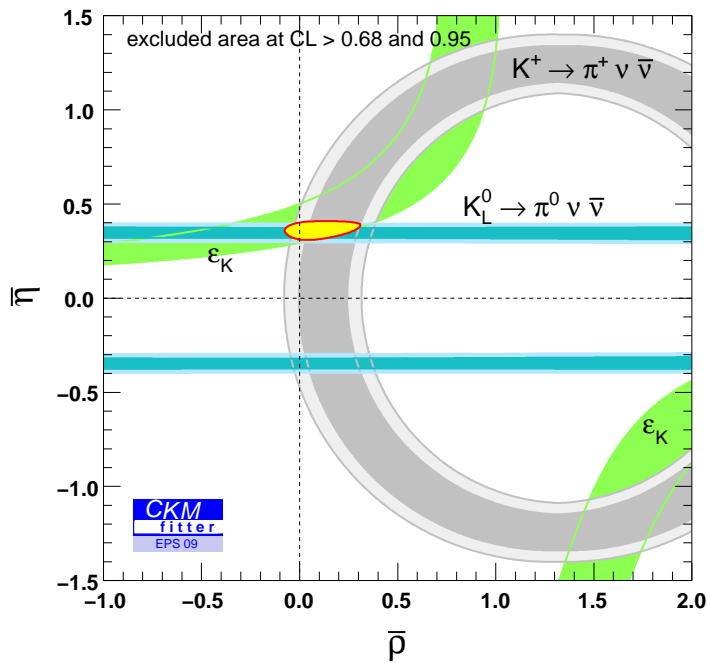}
}
\end{center}
\caption{A prediction of the possible constraint on the unitarity triangle made using the existing measurement 
   of $\epsilon_K$ and planned measurements of the branching rations of neutral and charged 
   $K\to \pi\nu\overline{\nu}$ decays.  This prediction has been made by the CKM Fitter 
   group~\cite{bevan:ckmfitter}.}
\label{bevan:fig:kaonckm}
\end{figure}

\section{\boldmath MEG: $\mu^\pm \to e^\pm \gamma$}

The lepton flavor violating process $\mu^\pm \to e^\pm \gamma$ has a branching fraction
of $\sim 10{-50}$~\cite{bevan:meg:smprediction}.  New physics could significantly enhance
the branching fraction of the signal to the current experimental limits of
{${\cal{O}}(10^{-11})$}~\cite{bevan:meg:currentlimit}.  The experimental challenge is to
identify a back to back $e^\pm$ and $\gamma$ from the decaying $\mu^\pm$.  This signal
process has to be isolated from irreducible physics backgrounds where
$\mu^\pm \to e^\pm \nu_\mu\overline{\nu}_e \gamma$ where the neutrinos are low energy, so that
the electron and photon are almost back-to-back in the final state.  The second main
background comes from $\mu^\pm \to e^\pm \nu_\mu\overline{\nu}_e$ decays with an in
time accidental photon.  As the final state neutrinos are go undetected by the experiment,
these two backgrounds are irreducible in the limit of low energy $\nu_\mu$ and
$\overline{\nu}_e$.  It is clearly desirable to have excellent tracking and calorimetry for
this experiment, and a high level of time resolution in order to reduce backgrounds
arising from accidental activity.

MEG has recorded data for approximately 3.4 million seconds during 2008, and will
resume taking data later this year.  The ultimate design goal is a single event sensitivity
of $1\e{-13}$ on $\br(\mu^\pm \to e^\pm \gamma)$.  During the first year of data taking
MEG encountered an issue with its drift chamber readout, that has since been corrected.
The data are currently being analyzed using a blind analysis technique, and the first results
from MEG are expected soon.   Several weeks after this conference the MEG collaboration 
released preliminary result based on their 2008 data sample.  This was
$\br(\mu^\pm \to e^\pm \gamma)< 3\e{-11}$~\cite{bevan:meg:lp}.  In comparison, the expected upper
limit from the 2009 run will be more than an order of magnitude lower~\cite{bevan:meg:recenttalk}.

\section{\boldmath $\mu \to e$ conversion experiments}

In addition to searching for LFV in the $\mu$ sector, it is possible to try and observe the behavior
of muonic-atoms, where a $\mu^-$ may spontaneously decay into an $e^-$ without the emission
of an associated neutrino.  The new physics scenarios that might lead to an observable 
signal can differ from the mechanisms giving enhancements to LFV in $\tau$ or $\mu$ decay.
These conversion experiments are very delicate and challenging enterprises,
and the best limit on $\mu \to e$ conversion obtained thus far is from the Sindrum-II 
experiment.  Studying gold the Sindrum-II collaboration placed an upper limit on 
$\mu \to e$ conversion at $<7\e{-13}$ (90\% C.L.)~\cite{bevan:mu2e:sindrum}.  There are two experimental programmes 
currently being planned: Mu2e~\cite{bevan:mu2e:mu2e} at Fermilab in the US, and 
COMET~\cite{bevan:mu2e:comet} at JPARC in Japan.  The
initial stages of both of these programmes would aim to start collecting data in 2013
and improve upon the Sindrum-II limits by a factor of 1000.

\section{Summary}

Flavor physics experiments have produced a mind-boggling array of measurements, and only
a few of the highlights of this experimental programme have been summarized in these proceedings.
The theoretical motivation for much of the programme has been to test the CKM paradigm, and this
has been shown to work exceptionally well in \B\ and \K\ decay.  Despite this good agreement,
there is still plenty of room for new physics effects to be manifest (and subsequently tested).
The tests of the CKM paradigm in charm decays is on the cusp of a new era as existing experiments having established charm mixing, are now
starting to probe for \CP\ violation in this sector.

There are a number of rare decay tests that are able to constrain new physics scenarios.  In some
of these cases, for example 2HDM constraints on \mhiggsp\ vs. \tanbeta\ using $\B^- \to \tau^- \nu$
decays, and LFV in $\tau$ decay, the constraints obtained from the \bfactories are more stringent 
than anything the LHC will be able to produce.
It may be possible for the LHC experiments to develop new techniques to marginally improve upon 
the existing constraints, however a Super Flavor Factory will be required to make a significant 
improvement over existing limits.  With regard to the related decay $\mu^\pm\to e^\pm\gamma$, the 
MEG experiment is working on finalizing a preliminary measurement from data recorded in 2008, 
and is expected to ultimately reach a single event sensitivity of $(30-50)\e{-13}$.  Significantly 
improved limits are expected from MEG using data taken in future runs (starting in the autumn of 2009).

The CLEO-c charm factory has produced a large number of branching fraction measurements, and using some of
these they have been able to measure form factors $f_D$ and $f_{D_s}$.  A detailed analysis of
the Dalitz Plot structure of $\Dz \to \Kz \pi^+\pi^-$ decays has been made, and this measurement will
lead to significant improvements in the model uncertainty of the model uncertainty on the $\gamma$
measurement using the GGSZ method.

The BES-III Charmonium factory has recorded 100 million $\psi(2S)$ decays and is in the process of
accumulating between 300 and 500 million \jpsi\ decays for precision measurements.

In addition to discovering many new particles, the \B-factories have been able to accumulate
data at the $\Upsilon(NS)$ resonance, where $N=1, \ldots 5$.  Using these data the have been
able to test lepton universality, search for light Higgs particles and place limits on light
dark matter scenarios.  KLOE and NA62 have also been able to test lepton universality through
precision measurements of \rk.

There are a number of planned flavor physics experiments to cover \B, \D, \K, $\tau$ and $\mu$
decays.  Together these measurements will provide a broad base of measurements to be performed
that will help to elucidate the detailed structure of any new physics found at the LHC.

\section{Acknowledgments}

The author would like to thank the meeting organizers for their invitation to this exhilarating
conference.  In addition I would also like to thank R. Harr, A. Schwartz, and J. G. Smith.
This work is funded by the UK Science and Technology Facilities Council.

\end{document}